# Photosensitive free-standing ultra-thin carbyne-gold films

Vlad Samyshkin[1], Anastasia Lelekova[1], Anton Osipov[1,2], Dmitrii Bukharov[1], Igor Skryabin[1], Sergey Arakelian[1], Stella Kutrovskaya,[1, 3,4]

[1] Department of Physics and Applied Mathematics, Stoletovs' Vladimir State University, Vladimir, Russia
[2] ILIT RAS — Branch of FSRC "Crystallography and Photonics" RAS, Moscow region, Russia
[3] School of Science, Westlake University, Zhejiang Province, China
[4] Institute of Natural Sciences, Westlake Institute for Advanced Study, Zhejiang Province, China

S.Kutrovskaya 11stella@mail.ru
A.Osipov osipov@vlsu.ru

**Abstract**
In the article we introduce the experiment of the photostimulation effect in the tunneling conductivity of free-standing thin C-Au films. We observe a sharp increase of the conductivity of hybrid film due to the electromagnetic exciting at the frequencies which are close to the plasmon resonance of gold nanoparticles. The use of carbyne threads as a stabilizing matrix makes it possible to obtain free-standing thin films that demonstrate a good structural stability. The tunnel current-voltage measurements demonstrate a strong dependence of the current value on the intensity of green laser radiation used to photostimulate thin C-Au in area of the measuring experiment.

**Keywords**
thin films, gold nanoparticle, carbyne, photocurrent

**Introduction**

The formation of ultrathin films containing metal nanoparticles is a promising task for applications in optoelectronics and photonics. Using such films one can significantly simplify the production of a wide class of optoelectronic devices which are based on transparent conducting metamaterials with flexible optical properties. Various methods of formation of such films based on the self-assembly of polymers, nanoparticles and molecules interacting at the air-water interface (Mueggenburg et al. 2007; He et al. 2010; He et al. 2011) or the liquid-liquid interface (Duan et al. 2004; Wang et al. 2006; Xia and Wang 2008) as well as modified Langmuir – Blodgett methods (Zasadzinski et al. 1994; Guo et al. 2003) and other methods have been developed recently.

The sacrificial layer method is one that is used most widely. In this method, the formation of individual nanostuctures occurs during the removal of the sacrificial layer situated below it (Jin et al. 2007; Wu et al. 2004; Xu et al. 2008). The sacrificial layer method is very suitable for fabrication of free-standing nanoparticles, fluorescent quantum dots, nanocomposites. (Jiang et al. 2004; Jiang, et al. 2004a; Jiang et al. 2006) However, it is not the universal one. One of its valuable alternatives is the film self-assembly method that allows to obtain a great variety of structures characterized by a reliable level of stability. In order to solve the stability problem for island like structures, one can combine colloidal nanoparticles with polymers in order to produce tailored hybrid materials (Vaia and Maguire 2007; Ofir et al. 2008; Kodiyan et al. 2012). This approach not only allows to increase the stability of colloidal particles, but also improved the functional characteristics of meta- and nano-materials for applications in sensors (Zamborini et al. 2002; Nam et al. 2004), microelectronics (Wei et al. 2012), and catalysis (Huh et al. 2010).

Nowadays, organic matrixes are extensively popular, for instance, the widespread materials such as the cellulose that contains a natural polysaccharide have been proposed for various applications in electronics (Fortunato et al. 2008). In addition, the usage of a carbon matrix for stabilization of metal nanoparticles and preventing their degradation in an environment appears to be an interesting and promising direction for the development of hybrid films with tailored optical properties (Bashouti et al. 2015). Here we shall focus specifically metal-carbon materials based on carbyne (Sladkov 1989). Carbyne is the fourth carbon allotrope that is composed by monoatomic linear chains of carbon. Carbyne is suitable for encapsulating the nanoobjects of an arbitrary shape and size, in particular, spherical nanoparticles of different metals. The distance between the atoms in a carbyne chain is of 0.133 nm for a single electron bond and 0.124 nm for a triple bond. Both these distances are less than the interatomic distance in graphite (0.142 nm). This is why, according to the theoretical estimates the robustness of carbyne based materials can be higher than one of graphene (Liu et al. 2013). Interestingly, carbyne crystals are predicted to be direct bandgap semiconductors featuring the gaps of 0.4 eV to 1.2 eV width depending on the electronic bond structure and the distortion level (Carbyne and carbynoid structures 2011).

Here we report on the experimental study of the photostimulation effect in the tunneling conductivity of free-standing thin C-Au films. Namely, we observe a sharp increase of the conductivity of this hybrid film when optically exciting the system at the frequencies close to the plasmon resonance in gold nanoparticles. We study thin films formed by spraying under atmospheric pressure of a colloidal system containing gold nanoparticles and carbon polymers (Kucherik et al. 2016). We deposited the solid constituent on an optically polished substrate for the morphology studies and on a special conductive grid for the transmission electron microscopy (TEM) studies. To measure the current-voltage characteristics as well as to examine the photosensitivity of our samples we used a scanning tunneling microscopy (STM). The characteristic features of tunneling current are nicely reproduced by our theoretical model that predicts the variation of the carrier mobility under the light action.

## The fabrication technic for free-standing C-Au films

The colloidal system containing metal nanoparticles and linear carbon chains – (carbyne) was prepared following the method described in (Kucherik et al. 2016). The average diameter of gold nanoparticles was about 10nm. To form a free-standing film we sputtered the colloidal solution through a nozzle of 0.3 mm diameter on a substrate placed within the distance of 10mm. We used a single shot spray regime to prepare a sufficiently thin film in order to be able to reveal its landscape by TEM. In contrast, the multishot spray regime was used to prepare the films suitable for the studies of conductivity. The overlap area covered by the sputtering process was up to 5 cm.

As shown in (Kutrovskaya et al. 2017), in the course of the sputtering process, a set of separately located droplets incorporating C-Au complexes is being formed on a substrate. These areas became growth points of islands of metal nanoparticles connected to each other by carbyne filaments that attach themselves to the substrate in the process of evaporation. We used the atomic-force microscopy (AFM), the Raman spectroscopy and the transmission electron technique for a comprehensive investigation of the deposited films.

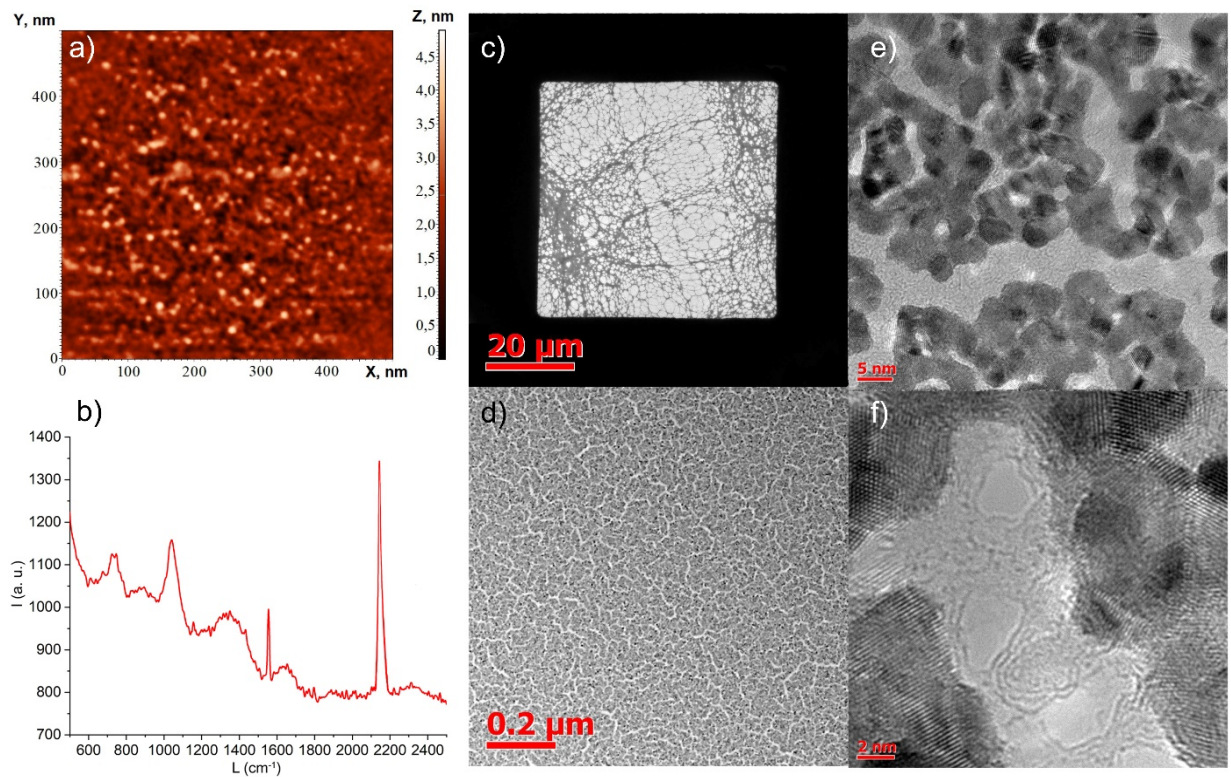

**Fig. 1** The deposited free-standing C-Au films. a) The AFM-image measured in the contact mode in vacuum by the Ntegra Aura device. b) The Raman spectra of a deposited layer. c) The characteristic high resolution TEM images of the studied film deposited on a copper grid. The film was formed by stretching inside the grid mesh. e),d)TEM-images taken with different magnifications that clearly show a fractal type of the obtained C-Au film. f) the high resolution TEM image that clearly shows carbyne based bridges connecting the neighboring gold nanoparticles

The Fig. 1 shows the experimental results obtained on the deposited gold-carbyne films. The AFM image allows to distinguish metal nanoparticles distributed within the stable inhomogeneous film. One can clearly see that the film is composed by several fractions. However, we detect no significant height variations (see the z-scale in Fig.1a). The Raman analysis shows that the formed carbon filaments are predominantly related to the polyyne allotrope of carbyne characterized by alternating single and triple electronic bonds. Indeed, the intense peaks at the wavelengths 2150cm$^{-1}$ and 1050cm$^{-1}$ correspond to triple and single electron bonds linking carbon atoms, respectively (see Fig.1b). The broadening of the peak corresponding to the C-C bond at 600sm$^{-1}$ is a result of deformation of the chains after deposition as well as of the Peierls distortion. The presence of hp2 - hybridization carbon allotrope is undoubtedly visible in the spectral vicinities of 1400 and 1650 sm$^{-1}$. These spectral features are characteristic of randomly oriented carbon threads.

For a detailed morphological analysis, we have performed the high resolution transmission electron microscopy (HR TEM) studies with a spatial resolution of up to 2 $\mathring{A}$. The set of characteristic magnified images shown in Fig.1(c-f) demonstrates the formation of free-standing thin films containing separate gold islands. Each island is formed by individual gold nanoparticles, while no uniform gold film is formed. The average diameter of the observed gold islands is about 30 nm and the distance between them is 5 nm (see Fig.1e). Carbon filaments form a random array connecting metal islands (see Fig.1f).

**Tunneling conductivity measurements**

To study the electrophysical properties of the deposited films we employed a set-up based on the Ntegra Aura device operating in the tunneling mode in vacuum. Fig. 2 schematically illustrates the experimental set-up we used. Firstly, we reconstruct the landscape of the film in order to be able to put the probe directly on the surface of the carbon matrix between metallic islands, as shown in the inset of Fig. 2a. The measured current-voltage curves are shown in Fig. 2b.

The black curve in Fig. 2b corresponds to the voltage-current dependence of the thin film placed on a pure aluminum substrate. The thickness of film was enough to neglect the substrate contribution into the collecting data of the tunneling current. It demonstrates a nearly linear behavior in the selected voltage range from -1 to 1 V. The presence of a nonzero current at zero voltage can be ascribed to the formation of an electron cloud near the metal surface. The I/V curve for a thin hybrid C-Au film looks similar to the black curve. It is also essentially linear. However, the value of the locking voltage is increased up to 0.27 V (see Fig. 2c) in the latter case as compared to the former case. We detect a significant increase of the tunneling current in the case of the laser illumination of the tunneling probe area.

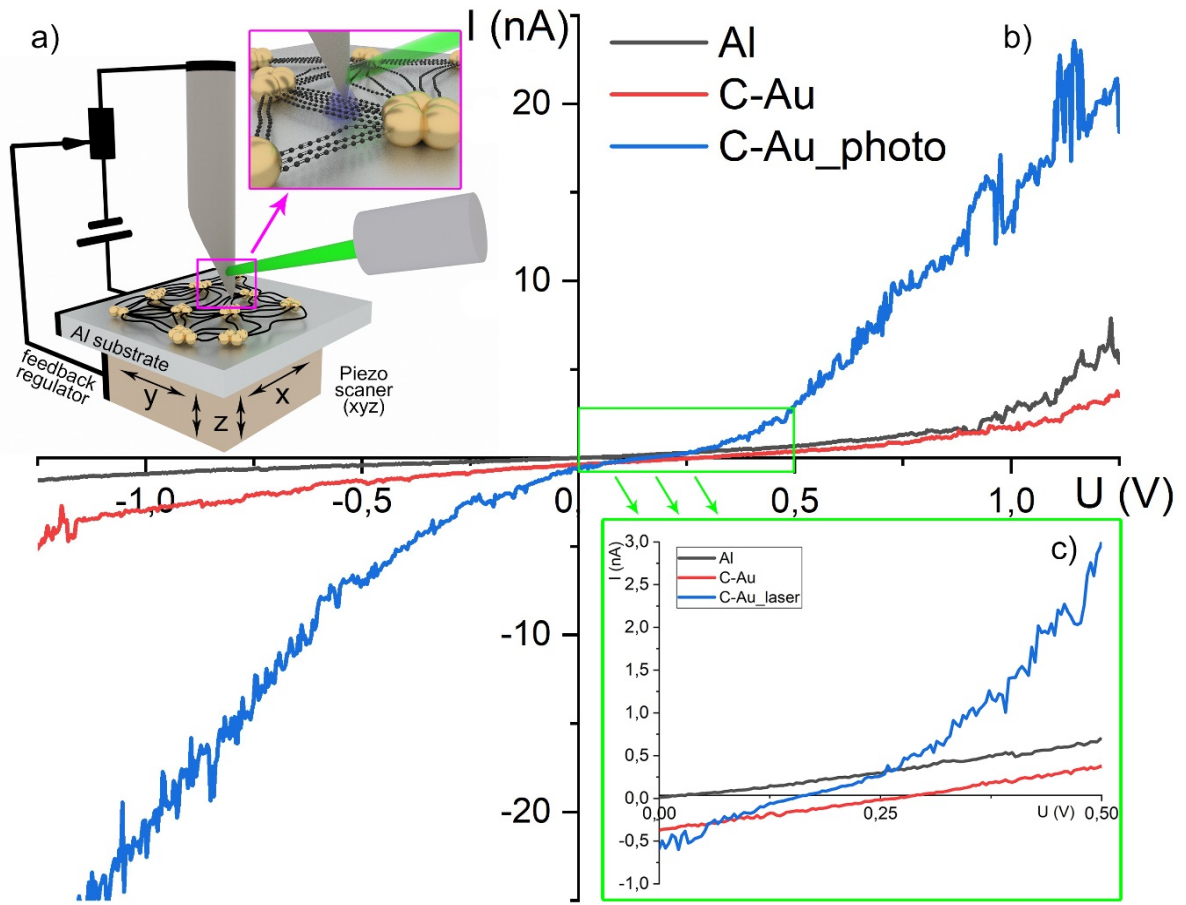

**Fig. 2** The tunneling current as a function of the voltage applied to the deposited free-standing C-Au films: a) schematically illustrates the layout of the main elements of the STM set-up. b) The current-voltage curves measured at various samples: the black line corresponds to the I/U curve of a pure aluminum substrate, the red curve corresponds to the C-Au thin film and the blue curve shows the I/U dependence of the C-Au thin film subjected to the excitation at the wavelength of 532 nm. c) The inset presents the behavior of the tunneling at a low applied voltage level

As a consequence, the I/V curve does not deviate from the linear behavior even in the small voltage region (see the blue curve in Fig.2b). The observed current amplification may be caused by the increase of the free charge carrier concentration due to the resonant absorption of the laser radiation of the 532 nm wavelength by gold nanoparticles.

The tunneling current spectroscopy results (see Fig.3) confirm this assumption. The increase of the distance between the STM probe and the thin film surface in the absence of laser irradiation results in a typical exponential decay of the tunneling current (see the black curve in the Fig.3). In contrast, the laser action leads to the appearance of steps in the current reduction curve that becomes nonmonotonous.

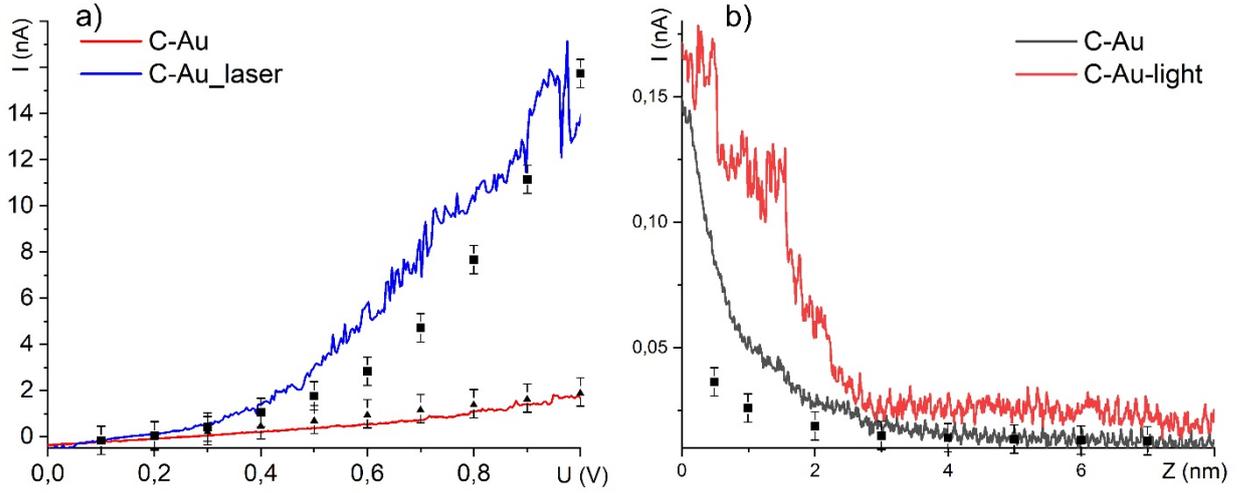

**Fig. 3** a) The positive branch of the experimentally measured current-voltage curve shown in Fig.2b compared to the predictions of the model marked by black points. b) The dependence of the tunneling current on the distance between the STM probe and the thin film surface: the red curve is measured under laser stimulation, the black curve is measured with the laser switched off. The results of the theoretical calculation are shown by black points

This can be interpreted as a manifestation of two effects: first, the decrease of the electron work function in gold nanoparticles due to the absorption of light (Wood 1981), second, the discontunous topological structure of the studied hybrid films. The conductivity in disordered metal-semiconductor structures is governed by the interplay between hopping and tunneling of electrons (Kavokin et al. 2017). Abrupt features in the current dependence on the distance between STM probe and the surface may be attributed to the thresholds between tunneling and hoping mechanisms of conductivity. Indeed, the electric current rapidly decreases at the distances of over than 2 nm that is in good agreement with the anticipated collapse of the tunneling conductivity at the distances of 2-3nm between islands (depending on the shape of the islands).

**Analytical model describing the current-voltage dependence of the tunnel current**

To describe the STM data, we have used the low voltage *(U)* model which takes into account only the electron tunneling between the conducting probe and the sample surface directly under the probe. The width of the tunnel junction is given by the distance between the probe and the sample surface. This way, the bias current *(I)* can be described as (Fishelson et al. 2017):

$$I(U) = A * U * e^{-k*d}, \tag{1}$$

where $k = \frac{4\pi\sqrt{2m\varphi}}{h}$ is the damping constant of the wave function in the potential barrier region,

m = 9,1*10-31 kg is the electron, φ= 6,5689*10-19 –J is the electron work function for gold, h = 6,6*10-34 J/s is the Planck constant, d is the probe – sample surface distance, A=$10^{-9}$ m is the scaling factor.

Ones the voltage increases and raises over *1V*, the tunneling current starts exhibiting a nonlinear growth as a function of the voltage (see Fig. 2) (Fishelson et al. 2017). The conductivity changes under the effect of light excitation of the deposited layer are likely to be caused by the change in the effective mobility of the carriers (Valverde-Aguilar et al. 2011). To account for this effect, we include the stationary photocurrent that emerges under the effect of laser irradiation into the model:

$$I = qG\Phi, \qquad (2)$$

where q is an electron charge, G is the photoelectric gain, Φ is the luminous flux density (Chen et al. 2011).

In its turn, the photoelectric gain can be defined as

$$G = \frac{\tau\mu U}{L^2}, \qquad (3)$$

where τ is the lifetime of photocarriers, μ is their mobility, U is the voltage applied to the sample, L is the distance between electrodes, for our case it is equal to the distance between the sample and the STM probe (Tanabe 2016).

According to this approach, the current-voltage curve was calculated for the low voltage range (from 0 to 1 V) using the following parameters τ=$10^{-6}$ s, L=0.1nm, f=$10^3$ W/sm$^2$, μ=0.41*$10^{-8}$ s. The calculation results are in a good agreement with the experimental data as Fig. 3 shows. One can clearly see that in the absence of the optical stimulation the current-voltage curve demonstrates the linear behavior at the low voltage region (see Fig. 3a). To reproduce the nonlinear dependence of the photocurrent, we have assumed the quadratic dependence of the carrier mobility on the light intensity that is typical for semiconductor crystals. Generally, the calculation results are in a good qualitative agreement with the experimental data. The present model can be applied to assess the electrical properties of the hybrid thin films containing metallic islands. The model calculation of the tunneling current shown in Fig. 3b correctly reproduces the decrease of the tunneling current with the increase of the distance between the STM probe and the surface. We note that the proposed approach only accounts for the carrier mobility directly in the measurement area of the tunneling current. The model neglects the electron hopping between metal islands (Kavokin et al. 2017), that leads to the supplementary variation of the carrier concentration. This effect will be a subject to our next studies.

## Conclusions

We have studied morphological and electrophysical properties of hybrid C-Au thin films. The use of carbyne threads as a stabilizing matrix makes it possible to obtain free-standing thin films that demonstrate a good structural stability. The tunnel current-voltage measurements demonstrate a strong dependence of the current value on the intensity of green laser light used to illuminate the sample. The observed phenomena are described within the theoretical model that ascribes the variations of the tunneling current to the variation of the carrier mobility. Clearly, the introduction of metal nanoparticles characterized by a variety of plasmon frequencies enables one to strongly improve the tunneling current for applications in nanoelectronic and photonic devices.

## Acknowledgements

This work was supported by RFBR grants 18-32-20006 mol_a_ved 19-32-90085_aspiranti and by Westlake University (Project No. 041020100118). This work was partially supported by the Ministry of Science and Higher Education and FSRC «Crystallography and Photonics» RAS.